\documentclass[aps,prd,preprint,tightenlines,superscriptaddress,nofootinbib,showpacs,byrevtex]{revtex4}
\usepackage{amssymb,latexsym}
\usepackage{amsmath,amsbsy}
\usepackage{epsfig,bm}
\usepackage{graphicx,comment}
\DeclareMathOperator{\tr}{tr}
\DeclareMathOperator{\Erfi}{Erf{}i}
\DeclareMathOperator{\Erf}{Erf}

\begin{document}
\def\a{{\alpha}}
\def\b{{\beta}}
\def\d{{\delta}}
\def\D{{\Delta}}
\def\e{{\varepsilon}}
\def\g{{\gamma}}
\def\G{{\Gamma}}
\def\k{{\kappa}}
\def\l{{\lambda}}
\def\L{{\Lambda}}
\def\m{{\mu}}
\def\n{{\nu}}
\def\o{{\omega}}
\def\O{{\Omega}}
\def\S{{\Sigma}}
\def\s{{\sigma}}
\def\th{{\theta}}

\def\ol#1{{\overline{#1}}}

\def\Dtslash{\tilde{D}\hskip-0.65em /}

\def\CPT{{$\chi$PT}}
\def\QCPT{{Q$\chi$PT}}
\def\PQCPT{{PQ$\chi$PT}}
\def\tr{\text{tr}}
\def\str{\text{str}}
\def\diag{\text{diag}}
\def\order{{\mathcal O}}

\def\cC{{\mathcal C}}
\def\cB{{\mathcal B}}
\def\cT{{\mathcal T}}
\def\cQ{{\mathcal Q}}
\def\cL{{\mathcal L}}
\def\cO{{\mathcal O}}
\def\cA{{\mathcal A}}
\def\cQ{{\mathcal Q}}
\def\cR{{\mathcal R}}
\def\cH{{\mathcal H}}
\def\cW{{\mathcal W}}
\def\cM{{\mathcal M}}
\def\St{{\tilde{\Sigma}}}
\def\S{{\Sigma}}
\def\qt{{\tilde{q}}}
\def\Dt{{\tilde{D}}}

\def\Slash#1{{#1\!\!\!\!\!\slash}}
\def\Dslash{D\!\!\!\!\slash}
\def\SppP{{\cal {P\!\!\!\!\hspace{0.04cm}\slash}}_\perp}
\def\ppslash{p^{\,\prime}\!\!\!\!\!\slash}
\def\nslash{n\!\!\!\slash}
\def\bnslash{\bar n\!\!\!\slash}
\def\pslash{p\!\!\!\slash}
\def\qslash{q\!\!\!\slash}
\def\lslash{l\!\!\!\slash}
\def\dslash{\partial\!\!\!\slash}
\def\Aslash{A\!\!\!\slash}
\def\OMIT#1{}

\newcommand{\CH}[2]{\chi_{#1,#2}}
\newcommand{\CHp}[3]{\chi_{#1,#2}^{#3}}
\newcommand{\bCH}[2]{\overline\chi_{#1,#2}}
\newcommand{\bCHp}[3]{\overline\chi_{#1,#2}^{#3}}

\newcommand{\nn}{\nonumber} 
\newcommand{\lc}{\lowercase}
\newcommand{\bn}{{\bar n}}
\newcommand{\bea}{\begin{eqnarray}}
\newcommand{\eea}{\end{eqnarray}}
\newcommand{\nb}{\bar n}
\newcommand{\bnp}{\bar n \!\cdot\! p}
\newcommand{\bnP}{\bar {\cal P}}
\newcommand{\ppP}{{\cal P}_\perp}
\newcommand{\bnPd}{\bar {\cal P}^{\raisebox{0.8mm}{\scriptsize$\dagger$}} }
\newcommand{\cP}{{\cal P}}
\newcommand{\cPslash}{ {\cal P}\!\!\!\!\slash}
\newcommand{\bs}{\!\hspace{0.05cm}}
\newcommand{\mcdot}{\!\cdot\!}
\newcommand{\Ub}{{\cal U}}
\newcommand{\cD}{{\cal D}}
\newcommand{\LQCD}{{\Lambda_{\rm QCD}}}
\newcommand{\np}{n \!\cdot\! p}

\newcommand{\mpsi}{M_\psi}
\newcommand{\qsqr}{Q^2}
\newcommand{\jpsi}{J/\psi}
\newcommand{\emax}{E_{max}}

\def\eqref#1{{(\ref{#1})}}

 
\title{Quarks with Twisted Boundary Conditions in the Epsilon Regime}

\author{Thomas Mehen}
\email[]{mehen@phy.duke.edu}
\affiliation{Department of Physics, Duke University, Durham, NC 27708, USA}
\affiliation{Jefferson Laboratory, 12000 Jefferson Ave., Newport News, VA 23606, USA}
\affiliation{Institute for Nuclear Theory, University of Washington, Seattle, WA 98195, USA}
\author{ Brian C.~Tiburzi}
\email[]{bctiburz@phy.duke.edu}
\affiliation{Department of Physics, Duke University, Durham, NC 27708, USA}
\affiliation{Institute for Nuclear Theory, University of Washington, Seattle, WA 98195, USA}
\date{\today}


\begin{abstract}

We study the effects of twisted boundary conditions on the quark fields in the epsilon regime of chiral perturbation theory.  
We consider the $SU(2)_L\times SU(2)_R$ chiral theory with non-degenerate quarks and the $SU(3)_L\times SU(3)_R$ chiral theory
with massless up and down quarks and massive strange quarks. The partition function and condensate  are derived for each
theory. Because flavor-neutral Goldstone bosons are unaffected by twisted boundary conditions chiral symmetry is still
restored in finite volumes. The dependence of the condensate on the twisting parameters can be used to extract the pion decay
constant from simulations in the epsilon regime. The relative contribution to the partition function from sectors of different
topological charge is numerically insensitive to twisted boundary conditions.

\end{abstract}

\maketitle

\section{Introduction}

Understanding the strongly interacting regime of Quantum Chromodynamics (QCD) remains a challenging problem in physics.
The dynamics is non-perturbative and results in the confinement of the colored quark and gluon degrees of freedom. 
Numerically simulating the gauge theory on a spacetime lattice provides a first principles method for 
determining the properties of the color neutral hadrons that are observed in nature. 
Recently, there has been much effort in applying effective field theory methods to quantitatively compute the
effect of lattice artifacts such as unphysical quark masses, quenching, finite lattice spacing and finite volume 
on lattice simulations of hadronic observables. Effective field theory calculations can be used to extrapolate the results of simulations
with unphysical parameters to real world observables~\cite{Bernard:2002yk}. 
One can also use the dependence on lattice artifacts to extract 
the parameters appearing in low energy effective Lagrangians of QCD.
 
Lattice QCD simulations usually employ periodic boundary conditions for the quark and gluon fields.
The results of a simulation should be independent of the choice of boundary conditions as one takes the volume to infinity,
so there is considerable freedom in choosing the boundary conditions for the fields.
However,  the action must be single valued so that observables are well-defined.
Fields need only be periodic up to a  transformation which is symmetry of the action. If a generic field 
$\phi$ depends on a lattice coordinate $x_i$ which is defined on the interval $0 < x_i < L$, then we must have
\bea 
\phi(x_i+L) = U \phi(x_i) \nn \, ,
\eea
where $U$ is a symmetry of the action and $U^\dagger U =1$. When $U \neq 1$, these are called twisted boundary conditions. 
This is equivalent to gauging a global symmetry of the 
action and then turning on a uniform background gauge field. In infinite space, a constant background gauge field can be removed by a gauge
transformation, so there are no physical consequences in the infinite volume limit. However,  
a uniform background gauge field can modify the physics of a compact space.
Different applications of twisted boundary conditions have appeared throughout the years~%
\cite{Gross:1982at,Roberge:1986mm,Wiese:1991ku,Luscher:1996sc,Bucarelli:1998mu,
Kiskis:2002gr,Guagnelli:2003hw,Kiskis:2003rd,Kim:2002np,Kim:2003xt}.%
\footnote{The use of twisted boundary conditions in the time direction can be utilized to 
produce an imaginary chemical potential~\cite{Dagotto:1989fw,Hasenfratz:1991ax,Alford:1998sd,
Hart:2000ef,Allton:2002zi,deForcrand:2002ci,D'Elia:2002gd,deForcrand:2003hx}.}
Recently, there has been a renewed interest in twisted boundary conditions in lattice QCD simulations %
\cite{Bedaque:2004kc,deDivitiis:2004kq,Sachrajda:2004mi,Bedaque:2004ax,Tiburzi:2005hg}.
This recent work has focused on using twisted boundary conditions on quarks   
as a means of avoiding the restriction of lattice momenta to integer multiples of $2 \pi/L$.

In this paper, we consider a different aspect of twisted boundary conditions.  Since dynamical effects due to twisting show up
as corrections at finite volume, it is natural to examine the effects of twisted boundary  conditions in the
$\varepsilon$-regime of chiral perturbation theory (\CPT), where volume effects are non-negligible.  \CPT\ is the low-energy
effective theory of QCD that encodes the dynamics of pseudo-Goldstone bosons that emerge from spontaneous breaking of chiral
symmetry. Spontaneous symmetry breaking does not occur in finite volume.  Therefore the perturbative expansion of \CPT\ must
breakdown in the limit $m_q \to 0$ with $L$ held constant. Indeed  when $m_\pi L \ll 1$ the zero momentum mode of the Goldstone
boson becomes strongly coupled. In this limit, the path integral over the zero modes must  be treated exactly, while nonzero
modes can be treated perturbatively. Computation of the chiral symmetry breaking order parameter in this regime shows that the
chiral condensate vanishes exponentially as $m_\pi^2 f_\pi^2 \beta V \to 0$, where $\beta V$ is the four-dimensional
volume of the lattice. For early references on these finite-size effects
in theories with spontaneously broken  symmetries,
see~\cite{Jolicoeur:1985iw,Leutwyler:1987ak,Gasser:1987ah,Gasser:1987zq,Hasenfratz:1989pk,Hansen:1990un,Hansen:1990yg}. For
later investigations and numerical simulations see, for example, 
\cite{Jackson:1996jb,Osborn:1998qb,Damgaard:1998xy,Damgaard:1999tk,Hernandez:1999cu,Damgaard:2001js,Damgaard:2002qe,Bietenholz:2003bj,Giusti:2004yp,Fukaya:2005yg}.

In the $p$-regime of chiral perturbation theory one takes $1/L \sim m_\pi \sim Q$ so propagators scale as $Q^{-2}$, vertices scale as $Q^2$ or
higher (since vertices have at least two derivatives or one insertion of the quark mass)  and loop integrations (or sums in finite volume) scale as
$Q^4$. The contribution to the partition function from an arbitrary vacuum graph scales as $Q^n$ where $n \geq 4 l - 2 I + 2 V = 2 l +2$, $l$ is the
number of loops, $I$ is the number of internal lines and $V$ is the number of vertices.  Thus, in this regime the partition function can be computed
in perturbation theory. In the $\varepsilon$-regime, $1/L \sim \varepsilon$ but $m_\pi \sim \varepsilon^2$. Thus, the zero momentum mode is singled
out because its propagator scales as $\varepsilon^{-2}$ while the nonzero mode propagators scale as $\varepsilon^{-1}$. The leading order
interactions of zero modes with themselves are proportional to $m_\pi^2 \sim \varepsilon^4$ so an arbitrary vacuum graph with only zero momentum
modes and leading order vertices scales as $\varepsilon^{4 l - 4I + 4 V} =\varepsilon^4$. Therefore, an infinite number of graphs with zero momentum
modes must be resummed at leading order. This resummation is straightforward   because the theory of zero modes alone is simply a matrix integral.

Now we see how this power counting is modified in the presence of twisted boundary conditions. 
The twisted boundary conditions induce a constant shift in the momentum of the propagators
of fields that are charged under the background gauge field:
\bea\label{prop}
\frac{1}{k_\mu^2+m_\pi^2} \to \frac{1}{(k_\mu + B_\mu)^2+m_\pi^2} \, ,
\eea
where $B_\mu$ is the background gauge field. For generic twists, $B_\mu \sim 1/L \sim \varepsilon$  and the propagator in Eq.~(\ref{prop}) scales as
$\varepsilon^{-1}$. The zero modes of fields charged under the background gauge field are no longer strongly coupled. However, if there are Goldstone
bosons that are not charged under the background  gauge field their zero modes will continue to be strongly coupled. In QCD with $N$ quarks, $B_\mu$
must couple either to baryon number~\cite{Bedaque:2004kc}, in which case all meson fields  are neutral, or an element of the $SU(N)$ flavor group. Because
quark mass differences and electric charge break the flavor symmetry down to the diagonal subgroup, we will only consider twists that are flavor diagonal.
Then the flavor-neutral Goldstone bosons will be neutral under the background field and there will still be a restoration  of chiral symmetry as $m_q
\to 0$ with $L$ fixed.~\footnote{Even for generic twists it is easy to show that there will be $N-1$ Goldstone bosons neutral under the
background gauge field.} The qualitative
behavior of the condensate as a function of $m_q$ is the same as without twisting. However, not all the zero modes of the coset $SU(N)_L \times
SU(N)_R/SU(N)_V$ are strongly coupled, only those valued in the diagonal subgroup. The behavior of the condensate is quantitatively different in this
case. For sufficiently small twists the background gauge field can be comparable in size to $m_\pi^2$ and the appropriate power counting is $B_\mu \sim
m_\pi \sim \varepsilon^2$. In this case the zero modes of all Goldstone bosons are strongly coupled. We will refer to the case of  $B_\mu \sim m_\pi
\sim \varepsilon^2$ as the twisted $\varepsilon$-regime and $B_\mu \sim \varepsilon, m_\pi \sim \varepsilon^2$ as the remnant $\varepsilon$-regime.

In the untwisted theory at lowest order in the $\e$ expansion,
the partition function depends only on the  mass term in the chiral Lagrangian.
Computation of the volume dependence of the chiral condensate can be used
to extract  $m_q \langle \ol q q \rangle_\infty$, where $\langle \ol q q \rangle_\infty$ is the infinite volume condensate. 
In the presence of twisted boundary
conditions, however, the kinetic term in the chiral Lagrangian gives a leading order
contribution to the partition function. By computing the finite volume condensate
as a function of both volume and twist, it is in principle possible to
extract both $m_q \langle \ol q q \rangle_\infty$ and $f_\pi$, where $f_\pi$ is the pion decay
constant. This latter quantity is usually extracted from the late time 
behavior of the axial correlator. Thus, twisted boundary
conditions give an alternative method by which one may extract $f_\pi$ from
simulations in the $\e$-regime.

We also evaluate the projection of the partition function onto sectors with fixed topological charge by including a $\theta$-term in the chiral Lagrangian and Fourier
transforming with respect to the $\theta$ parameter. This calculation can be done analytically for the untwisted theory ~\cite{Leutwyler:1992yt}. We perform  the
calculation for the isospin symmetric $SU(2)_L \times SU(2)_R$ chiral  theory. Twisted boundary conditions do not dramatically change the results of
Ref.~\cite{Leutwyler:1992yt}. In particular, the observation that the QCD partition function is dominated by the sector with zero topological charge when $m_q \langle \ol q q
\rangle_\infty \beta V < 1$ still holds in the presence of twisted boundary conditions.

The paper is organized as follows: in Section~\ref{s:gen}, we describe in detail the twisted boundary conditions imposed on the quark fields in QCD
with $N$ quark flavors and the twisted finite-volume meson propagator. We write down a general formula for the  partition function in the presence of
twisted boundary conditions. In Sections~\ref{s:SU2} and \ref{s:SU3}, we evaluate this partition function  as well as the finite volume condensate for
twisted $SU(2)_L\times SU(2)_R$ and $SU(3)_L\times SU(3)_R$ chiral theories, respectively. We consider both the
twisted $\varepsilon$-regime and the remnant $\varepsilon$-regime. In
Section~\ref{s:pert}, we compute the projection of the partition function of the $SU(2)_L\times SU(2)_R$ chiral theory onto sectors of fixed
topological charge. Section~\ref{s:summy} contains a brief summary and  discussion of directions for future work.

\section{Generalities} \label{s:gen}

Let us consider lattice QCD gauge theory with $N$ quark flavors. In a finite volume, the boundary conditions on the quark fields must be chosen so
that action is single-valued. Thus the generator of any flavor symmetry of the action  can be used to specify the boundary conditions of the quark
fields. In the $i^{\text{th}}$ spatial direction,  we impose flavor diagonal twisted boundary conditions of the form
\begin{equation}
q(x_i + L ) = e^{i \Theta_i} q(x_i),
\end{equation}
where the Abelian phases $\Theta_\mu$ can be expanded in the Cartan subalgebra of $U(N)$.  Instead of using the Cartan generator basis,  we  will
sometimes choose to write the decomposition of twists in the flavor basis: $(\Theta_\mu)_{ab} = \d_{ab} \theta_\mu^a$, where $\theta_4^a = 0$ for
all flavors. Furthermore due to periodicity,  we need only consider twist angles $\th_i^a$ in the range $-\pi < \th_i^a \leq \pi$.

It is convenient to introduce quark fields $\qt(x)$ defined by
\begin{equation}
\qt (x) = U^\dagger(x) q(x),
\end{equation}
with
\begin{equation}
U(x) = e^{i \Theta_\mu  x_\mu / L}
.\end{equation}
The $\tilde q(x)$ fields satisfy periodic boundary conditions. 
The quark part of the QCD Lagrangian can then be rewritten in the form
\begin{equation}
\cL = \ol \qt ( \Dtslash + M ) \qt,
\end{equation}
where the effect of twisting appears as a background gauge field
\begin{equation}
\Dt_\mu = \partial_\mu + i B_\mu, 
\end{equation}
where $B_\mu = \Theta_\mu / L$.

To describe QCD at low energies, one turns to an effective field theory, chiral perturbation theory, 
that is written in terms of the pseudo-Goldstone bosons that appear from chiral symmetry breaking. 
As shown in \cite{Sachrajda:2004mi}, in the presence of twisting, this low energy effective theory of QCD
can be rewritten in terms of a modified coset field $\St(x)$ given by 
\begin{equation}
\St(x) = U^\dagger (x) \S(x) U (x),
\end{equation}
that satisfies periodic boundary conditions. The coset field $\St(x)$, however, 
has non-vanishing charge in the background gauge field, $B_\mu$. 
In terms of the modified coset field, the chiral Lagrangian appears as
\begin{equation} \label{eq:L}
\cL = \frac{f_\pi^2}{8} 
\left[
\tr (\Dt_\mu \St \Dt_\mu \St^\dagger)
-
\tr (\St \chi + \chi \St^\dagger)
\right],
\end{equation}
where $\chi$ is proportional to the quark mass and the covariant derivative  is  
\begin{equation}
\Dt_\mu \St = \partial_\mu \St + i [ B_\mu, \St].  
\end{equation}

From the Lagrangian in Eq.~\eqref{eq:L}, we can deduce the propagator of the meson fields.  
For our general discussion, we begin by omitting the flavor structure of the propagator from our consideration. 
In a finite volume with dimensions $L \times L \times L \times \beta$, the meson propagator for the mode $n_\mu = (\bm{n}, n_4)$ is proportional to
\begin{equation}
\frac{1}{(2 \pi n_4/ \b )^2 + (2 \pi \mathbf{n} + \mathbf{\Theta})^2/L^2 + m_\pi^2}
,\end{equation}
where $\bm{n} \in \mathbb{Z}^3$ and $n_4 \in \mathbb{Z}$ due to the periodic boundary conditions and 
the $\bm{\Theta}$ term arises from the background gauge field. For power counting we will always 
assume $\beta \sim 1/L \sim \epsilon$. When one works in the $\varepsilon$-regime, where $m_\pi L \ll 1$,
the zero mode $(n_\mu = 0)$ of the finite-volume propagator ordinarily is strongly coupled.
Twisting however has introduced an effective mass squared $B_\mu^2 + m_\pi^2$
for the zero mode. Because the flavor-neutral mesons are unaffected by the twisting, 
there are always strongly coupled zero modes in the group manifold. 
In the twisted $\varepsilon$-regime all Goldstone boson zero modes are strongly coupled.
Since the zero modes must be treated non-perturbatively, we separate the field into two pieces
\begin{equation} \label{eq:field}
\St(x) = \St_0 e^{2 i \tilde \phi (x)/f_\pi},
\end{equation}
where $\St_0$ is the zero mode and the $\tilde{\phi}$ contains the non-zero modes of the pion fields.

Returning to the Lagrangian in Eq.~\eqref{eq:L}, 
we see that to lowest order the partition function in the twisted $\varepsilon$-regime is
\begin{equation} \label{eq:integral}
Z = 
\int \cD \St_0 \, 
\exp 
\left\{ 
\frac{\beta V f_\pi^2}{4} 
\tr [B_\mu \St_0 B_\mu \St_0^\dagger 
+  
\Re \text{e} (\chi \St_0 ) ]
\right\} \, ,\end{equation}
where $V = L^3$. We have dropped a $B_\mu$ dependent constant which only modifies the normalization of the partition function  and is physically irrelevant. Note that
this partition function is very similar in form to that which arises when considering QCD with an isospin chemical
potential~\cite{Toublan:1999hx,Splittorff:2003cu,Akemann:2004dr}. However, an isospin chemical potential induces a $B_\mu$ that is time-like rather than
space-like and the contribution to the action comes with the opposite sign. For our treatment of the zero modes in the $SU(2)_L \times SU(2)_R$ and
$SU(3)_L \times SU(3)_R$ chiral theories, the expression for the partition  function in Eq.~\eqref{eq:integral} will be our starting point.  We will
also consider the partition function for the remnant $\varepsilon$-regime, where only the flavor-neutral mesons have strongly  coupled zero modes. In
this case, the matrix $\St_0$ belongs to the diagonal sub-manifold and only the quark mass term survives in Eq.~\eqref{eq:integral}.

\section{Twisted $\e$-regime of $SU(2)$} \label{s:SU2}

In this section we consider the effects of twisted boundary conditions on the quark fields in two 
flavor \CPT\ in the twisted $\e$-regime. The simplicity of the group manifold will allow us to 
reduce the zero mode contribution to the partition function to a one-dimensional 
integral which is easily evaluated numerically. The dependence of the finite-volume quark condensate on twisting angle is then investigated. 
We also derive the partition function in the remnant $\epsilon$ regime when only the $\pi^0$ zero mode
is strongly coupled.  

For two quark flavors, the general form of the background gauge field is
\begin{equation} \label{eq:charge}
B_\mu = \diag ( B_\mu^u, B_\mu^d ) = \frac{1}{2} (B^u_\mu + B^d_\mu)  + \frac{1}{2} (B^u_\mu - B^d_\mu) \sigma_3
.\end{equation}
The matrix integral appearing in Eq.~\eqref{eq:integral} 
can largely be carried out due to the simple structure of the $SU(2)$ group manifold. 
First notice the term proportional to the identity in Eq.~\eqref{eq:charge} decouples 
from the zero mode integration. To perform the remaining integration in the twisted $\e$-regime,
we use a familiar parametrization of $SU(2)$
\begin{equation}
\St_0 = \cos \a + i \hat{\bm{n}} \cdot \bm{\sigma} \sin \a 
,\end{equation}
where $\hat{\bm{n}}$ is a three-dimensional unit vector. 
With this choice for the arbitrary element of $SU(2)$, the group measure, normalized to unity, is
\begin{equation}
\int \cD \St_0 
= 
\frac{1}{2 \pi^2} \int d\Omega_{\hat{\bm{n}}} \int_0^{\pi} d\a \, \sin^2 \a
,\end{equation}
and the partition function is straightforwardly evaluated.

The final result can be cast into 
the form of a one-dimensional integral over well known functions. 
To this end, we define the dimensionless variables $s$ and $t$ 
\begin{eqnarray} \label{eqn:s}
s &=&  \frac{1}{8} f_\pi^2 \b V \, \tr (\chi) = \frac{1}{4} f_\pi^2 \b V m_\pi^2 =  -\frac{m_u+m_d}{2} \langle \bar q q \rangle_{\infty} \b V , \\
t &=&  \frac{1}{4} (\bm{B}^u - \bm{B}^d)^2 f_\pi^2 \b V = \frac{(\bm{\theta}^u - \bm{\theta}^d)^2}{4 L^2} f_\pi^2 \b V  
\label{eqn:t},\end{eqnarray}
where $m_\pi$ is the pion mass in the infinite volume limit.
The variable $s$ is the ordinary parameter relevant to the $\e$-regime \cite{Gasser:1987ah}, 
while the variable $t$ contains the new effects due to twisting the quark fields at the boundary of the lattice.  
Notice $t$ is zero only if in each spatial direction the quark phases at the boundary are pairwise identical. 
Furthermore in evaluating the matrix integral, we do not need to assume that the mass matrix is proportional 
to the identity. All results at this order in the chiral expansion depend only on $\tr (\chi)$. 
Terms in $\cL^{(4)}$ lead to a departure from this result, but these are beyond the order we work.

The angular integral in the group measure can be done analytically. In terms of the variables $t$ and $s$, 
the partition function to leading order in the twisted $\e$-regime is
\begin{equation} \label{eq:result}
Z(s,t) 
= \frac{1}{\sqrt{\pi}} \int_0^{\pi} d\a \,
\frac{\sin \a}{\sqrt{t}} 
\Erfi ( \sqrt{t} \sin \a )
\exp (2 \, s \cos \a - t \sin^2 \a)
,\end{equation}
where $\Erfi(x) = \Erf(ix)/i$, and  $\Erf(x)$ is the standard error function.
Untwisting the quark fields, i.e.~for $t = 0$, we arrive at
\begin{equation}
Z(s,0) = \frac{2}{\pi} \int_0^{\pi}
d\a \,
\exp ( 2 \, s \cos \a) \sin^2 \a
,\end{equation}
which is the form of the ordinary $SU(2)$ group integral in the $\e$-regime~\cite{Gasser:1987ah}.
In terms of a modified Bessel function, we have $Z(s,0) = I_1(2 s)/ s$.

In the twisted $\e$-regime, $B_\mu \sim m_\pi \sim \e^2$ and $t/s \sim 1$,
but in the remnant $\e$-regime
\begin{equation}
\frac{t}{s} \propto \frac{B_\mu^2}{m_\pi^2}  \sim \frac{1}{\varepsilon^2} \, .
\end{equation}
So the large $t$ limit corresponds to the remnant $\e$-regime. Taking the asymptotic limit of the error function,
\begin{equation}
\lim_{x \to \infty} \frac{\Erfi \sqrt{x}}{\sqrt{x}} = \frac{e^x}{\sqrt{\pi} x}
,\end{equation} 
we find the $t$-dependence in Eq.~\eqref{eq:result} reduces to an overall multiplicative factor which can be pulled outside the integral.
In general, one is interested in computing expectation values of operators which can be calculated by
adding sources to the Lagrangian and then differentiating the logarithm of the partition function with respect to these sources. 
The condensate is obtained by differentiating log $Z(s,t)$ with respect to $s$ (see Eq.~(\ref{cond}) below).
So the multiplicative dependence on $t$ is a trivial modification of the normalization of the path integral which
can be ignored. Thus, for large $t$, the partition function is
\begin{equation} \label{eq:rem2}
Z(s) \propto  \int_0^{\pi} d\a \, \exp ( 2 \, s \cos \a )  \propto I_0(2 s) \, .
\end{equation}
The $\pi^0$ zero mode has a mass 
$\sim m_\pi \sim \e^2$ but the charged pions have an effective mass $\sim | B_\mu | \sim \e$.
Physically, one expects to be able to integrate out the charged pion zero modes in this
limit. The twisted $\e$-regime partition function  should reduce 
to that of the remnant $\varepsilon$-regime, in which only the  $\pi^0$ zero mode is dynamical.
The remnant $\e$-regime partition function is obtained using the parametrization
\begin{equation}
\St_0 = \exp ( i \a  \, \sigma_3 ),
\end{equation}
and the group integral measure is trivial since it is an integral over a $U(1)$ subgroup
of the coset $SU(2)_L \times SU(2)_R/SU_V(2)$. This immediately leads to a partition function 
identical to that in Eq.~\eqref{eq:rem2}.

\begin{figure}
\epsfig{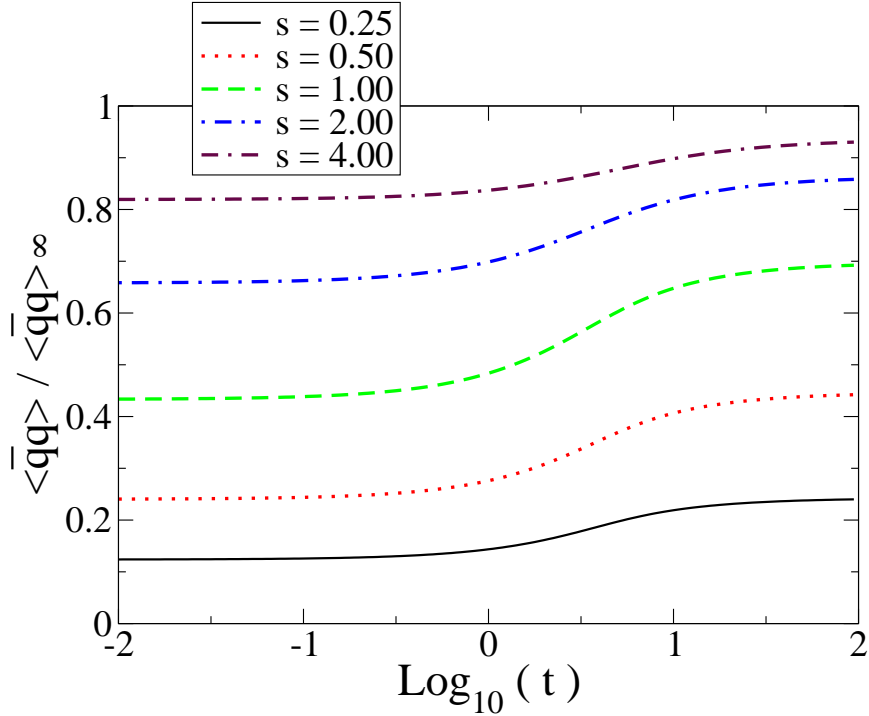}
\caption{The quark condensate of the $SU(2)_L \times SU(2)_R$ theory with twisted boundary conditions is plotted as a function of the 
twisting parameter $t$, for several values of $s$.}
\label{f:tSU2}
\end{figure}

The quark condensates for the up and down quarks are identical, $\langle \ol u u \rangle = \langle \ol d d \rangle$,
because the partition function depends only on the trace of the mass matrix $\tr (\chi)$. 
We denote either of the condensates by $\langle \ol q q \rangle$. They are calculated by
differentiating the partition function
\begin{equation}\label{cond}
\frac{\langle \ol q q \rangle \phantom{88}}{\langle \ol q q \rangle_\infty} = \frac{1}{2} \, \frac{\partial}{\partial s} \log Z(s,t)
,\end{equation}
\begin{figure}
\epsfig{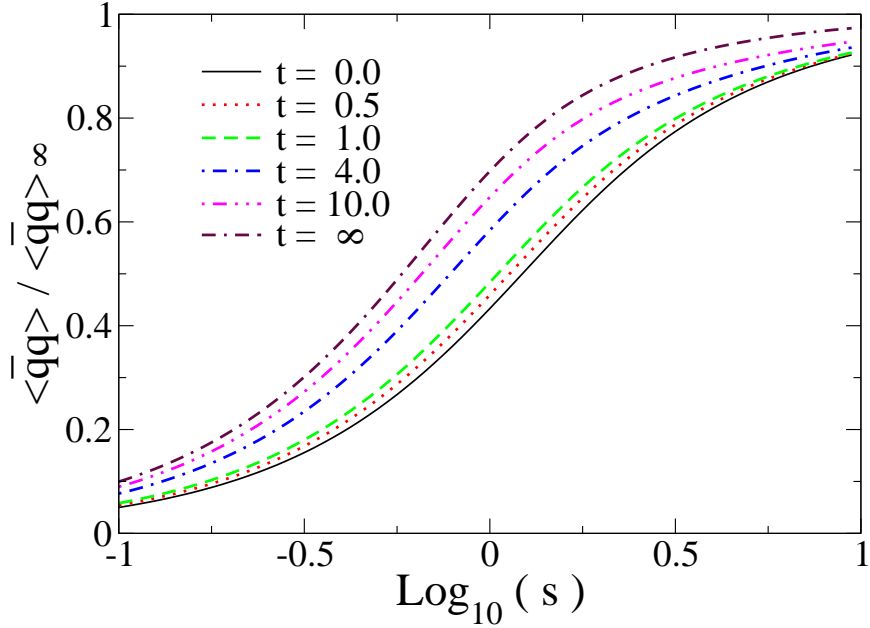}
\caption{The quark condensate of the $SU(2)_L \times SU(2)_R$  theory with twisted boundary conditions is plotted as a function of $s$, 
for several values of the twisting parameter $t$. Results in the remnant $\e$-regime correspond to the curve denoted $t = \infty$.}
\label{f:SU2}
\end{figure}
In Figure~\ref{f:SU2}, we show the dependence of the quark condensate on the twisting parameter $t$. For $t=0$, the result
is just that of the ordinary $\e$-regime of $SU(2)$. The condensate decreases with $s$, and vanishes as $s\to 0$ due to the
restoration of chiral symmetry at finite volume. For $t \neq 0$ there are twisted boundary conditions on the quark fields
and the qualitative behavior of the condensate as a function of $s$ is unchanged. The value of the finite-volume
condensate, however, is altered and depends on the twisting variable $t$. As $t$ becomes larger, the  effective mass of the
charged pion zero modes increases due to the twisting and these modes can no longer conspire to restore chiral symmetry.
Therefore the condensate increases in magnitude with $t$ for fixed $s$. When $t/s$ is very large, the theory is in the
remnant $\e$-regime and only the $\pi^0$  zero mode is relevant. The result for the remnant $\e$-regime is denoted by
the curve $t = \infty$.  We see that the behavior of the condensate for non-zero $t$  approaches that of the remnant
$\e$-regime in the limit of large $t$.  In Figure~\ref{f:tSU2} we plot the dependence of the quark condensate as a function
of $t$ for fixed $s$. Mapping out the dependence of the finite-volume condensate as a function of $\bm{\th^u}-\bm{\th^d}$
for fixed four-volume and $s$ gives one an alternate method to determine $f_\pi$. As $s \to 0$ the condensate must vanish
because of the restoration of chiral symmetry and therefore the value of the condensate is insensitive to the twisted
boundary  conditions. As $s \to \infty$ one is approaching the infinite volume limit and all observables should be 
insensitive to boundary conditions. The condensate is most sensitive to twisted boundary conditions near $s=1$.

\section{Twisted $\e$-regime of $SU(3)$} \label{s:SU3}

In this section, we investigate the consequences of twisted 
boundary conditions on QCD with three flavors of quark fields. 
The group integration over the $SU(3)$ manifold is more complicated than $SU(2)$. Our formulae
for the matrix integral in the most general case is an eight dimensional integral which 
we do not know how to perform analytically. To simplify numerical analysis 
of the partition function, we work in the limit of massless up and down 
quarks and choose isospin symmetric boundary conditions. The partition function in this 
case can be reduced to a one-dimensional integral and the dependence of the strangeness condensate 
on twisting angles and strange quark mass can be easily calculated. 


For the case of the $SU(3)$ flavor group, the flavor-diagonal  background gauge field is
\begin{equation}
B_\mu = \diag (B_\mu^u, B_\mu^d, B_\mu^s ) = B^0_\mu + B^3_\mu \l_3 + B^8_\mu \l_8
,\end{equation}
where $B^0_\mu = \frac{1}{3} (B_\mu^u + B_\mu^d + B_\mu^s )$,
$B^3_\mu = \frac{1}{2}(B_\mu^u -  B_\mu^d)$, and 
$B^8_\mu = \frac{1}{2 \sqrt{3}}(B_\mu^u + B_\mu^d - 2 B_\mu^s )$.
In Eq.~\eqref{eq:integral}, any terms involving $B^0_\mu$ decouple from the group integral.   
For the arbitrary group element of $SU(3)$, we use a convenient form~\cite{Byrd:1998uw,Byrd:1998??} 
that is a generalization of the Euler angle decomposition,
\begin{equation}
\St_0 
= 
e^{- i \l_3 \a} e^{i \l_2 \b} e^{- i \l_3 \g} e^{i \l_5 \psi}
e^{- i \l_3 a} e^{i \l_2 b} e^{- i \l_3 c} e^{- i \l_8 \phi}
.\end{equation}
With this decomposition, the invariant volume element of the $SU(3)$ group is~\cite{Byrd:1998uw,Byrd:1998??,Holland:1969??}
\begin{equation}
\int \cD \St_0 
\propto 
\int_0^{\sqrt{3}\pi} d\phi
\int_0^\pi d\a\, d\g \, da \, dc  
\int_0^{\pi/2} d\b \, db \, d\psi \,
\sin 2 \b \,
\sin 2 b \, 
\sin 2 \psi \, 
\sin^2 \psi
\, .\end{equation} 
(Note that in this case we have not chosen to normalize the group integral to 1.)

For the moment, let us consider three independent flavor twists and three non-degenerate
quark masses. In this general case, we can write the partition function in the twisted $\e$-regime as a function of the quark mass dependent variables
$s^a = \frac{1}{4} \chi^a f_\pi^2 \b V$ and the various combinations of flavor twists, $t^{ab} = \frac{1}{4} \bm{B}^a \cdot \bm{B}^b f_\pi^2 \b V$. 
We find the partition function has the form
\begin{equation}\label{Z3}
Z(s^a , t^{ab} )  = N \int \cD \St_0
\exp[f(s^u, s^d, s^s) + g(t^{33},t^{38},t^{88})]
,\end{equation}
where the quark mass dependent term in Eq.~\eqref{eq:integral} contributes to the function $f(s^u, s^d, s^s)$, which is given by
\begin{eqnarray}
f(s^u, s^d, s^s) 
&=&   
s^s \cos \psi \cos \frac{2 \phi}{\sqrt{3}}  \\
&& \hspace{-0.5 in}+\left[ 
s^d \cos 
\left( 
a + c + \a + \g - \frac{\phi}{\sqrt{3}}
\right)
+ 
s^u \cos \psi
\cos 
\left( 
a + c + \a + \g + \frac{\phi}{\sqrt{3}}
\right)
\right] \cos b \cos \b \notag \\
&&\hspace{-0.5 in} - 
\left[ 
s^u \cos 
\left( 
a - c - \a + \g - \frac{\phi}{\sqrt{3}}
\right)
+ 
s^d \cos \psi
\cos 
\left( 
a - c - \a + \g + \frac{\phi}{\sqrt{3}}
\right)
\right] \sin b \sin \b \notag 
,\end{eqnarray}
and the term in Eq.~\eqref{eq:integral} which depends on the background gauge field contributes to the function $g(t^{33}, t^{38}, t^{88})$, which is given by
\begin{eqnarray}
g(t^{33},t^{38},t^{88}) &=& \frac{1}{2} \, t^{33} \left[ ( 3 + \cos 2 \psi) \cos 2 \b \cos 2 b - 4 \cos ( 2 a + 2 \g ) \cos \psi \sin 2 \b \sin 2 b \right]
\notag \\
&&  
- \sqrt{3} \, t^{38} ( \cos 2 \b + \cos 2 b ) \sin^2 \psi
+ \frac{3}{2} \, t^{88} \cos 2 \psi
.\end{eqnarray} 
The full expression for the partition function in Eq.~(\ref{Z3}) is an eight dimensional integral which appears difficult
to evaluate analytically in the most general case. We will make two assumptions which greatly simplify 
the evaluation of the partition function. It is convenient to choose isospin symmetric boundary conditions 
with $B^d_\mu = B^u_\mu$. In this case, we have
\begin{equation}
g(t') = \frac{t'}{2} \cos 2 \psi
,\end{equation}
where $t' = \frac{1}{4} (\bm{B}^u - \bm{B}^s)^2 f_\pi^2 \b V$. Secondly, we make the physically relevant approximation
of neglecting up and down quarks masses while letting the strange quark mass be nonzero. Then the action only depends 
on the angles $\phi$ and $\psi$ and all but one integral can be done analytically:
\begin{eqnarray}
Z(s^s , t' ) 
&=& 
N' \int_0^{\sqrt{3} \pi} d\phi \int_0^{\pi/2} d\psi  \, \sin 2 \psi \, \sin^2 \psi \, 
\exp \left( s^s \cos \psi \, \cos \frac{2 \phi}{\sqrt{3}} + \frac{t'}{2} \cos 2 \psi \right) \notag \\
&=&
N'' \int_0^{1} dx  \, x (1 - x^2) 
I_0 (s^s x) \, e^{ t' (x^2 -1/2)} 
\label{eq:result2}
,\end{eqnarray}
where $I_0(x)$ is a modified Bessel function.

The strangeness condensate, $\langle \ol s s \rangle$, is given by
\begin{equation} \label{eq:sc}
\frac{\langle \ol s s \rangle \phantom{88}}{\langle \ol s s \rangle_\infty} = \frac{\partial}{\partial s^s} \log Z(s^s,t')
,\end{equation}
where the corresponding condensate in the infinite volume limit is denoted by 
$\langle \ol s s \rangle_\infty$.
\begin{figure}
\epsfig{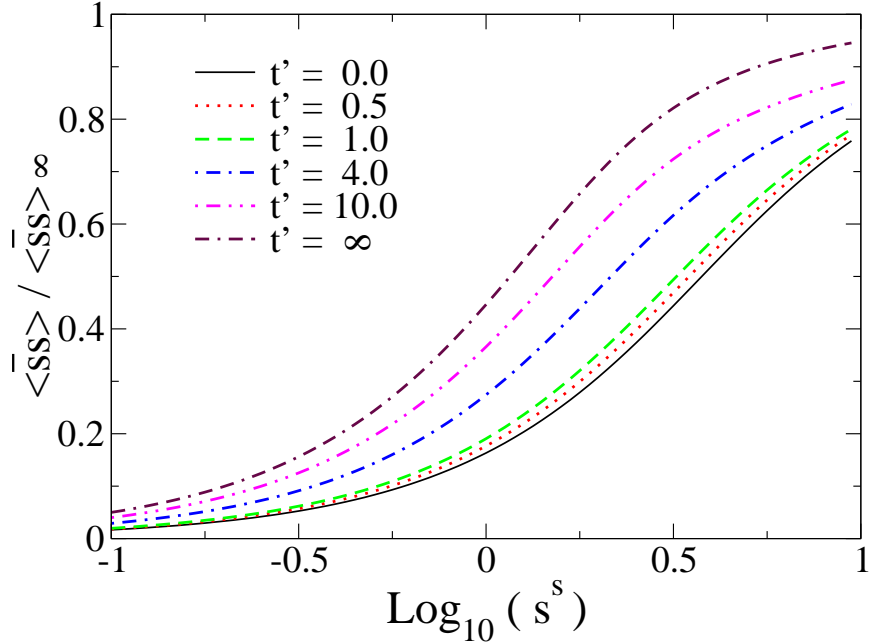}
\caption{The strangeness condensate in the $SU(3)_L \times SU(3)_R$ theory with twisted boundary conditions
 as a function of  $s^s$, for several values of the twisting parameter $t'$.
The remnant $\e$-regime corresponds to the curve denoted $t' = \infty$. Note that $\bm{B_u}=\bm{B_d}$ and $m_u = m_d = 0$.}
\label{f:SU3}
\end{figure}
In Figure~\ref{f:SU3}, we show the dependence of the strangeness condensate
on the twisting parameter $t'$. The result for $t'=0$, corresponding to
the case of periodic boundary conditions, is the finite-volume behavior of the condensate
in the ordinary $\e$-regime. Like the $\langle \bar q q \rangle$ condensate in the $SU(2)_L \times SU(2)_R$
theory,  the $\langle \bar s s \rangle$ condensate decreases monotonically with $m_s$.
When $t'$ is non-zero, the theory is in the twisted $\e$-regime and the qualitative behavior as a function of $s^s$ is unchanged. 
However, the flavor non-diagonal zero modes become less important than the flavor diagonal modes. 
The condensate is larger in the presence of twisted boundary conditions. 
The greatest enhancement is achieved when $t'/s^s \gg 1$, which is the remnant $\e$-regime.

Although we have not been able to analytically derive the large $t'$ limit of the strangeness condensate
directly from Eqs.~\eqref{eq:result2} and \eqref{eq:sc}, we can treat the remnant $\e$-regime exactly
when $m_u = m_d =0$. In the remnant $\e$-regime, only the $\pi^0$ and $\eta$ zero modes remain strongly coupled.
The integral over the $SU(3)_L\times SU(3)_R/SU(3)_V$ coset reduces to integrals over
the $U(1)$ diagonal subgroups. Therefore, the measure is trivial and $\St_0$ can be parametrized 
by the diagonal matrix $\St_0=\rm{diag}(e^{i \theta_1},e^{i \theta_2},e^{-i(\theta_1+\theta_2)})$.
Thus, the partition function in the remnant $\e$-regime is
\begin{equation}
Z(s^a) = N \int_0^{2 \pi} d \th_1 \int_0^{ 2 \pi} d \th_2 
\exp \left[ s^u \cos \th_1 + s^d \cos \th_2 + s^s \cos (\th_1 + \th_2)  \right]
,\end{equation}
To compare with our numerical results in the twisted $\e$-regime, 
we set $s^u = s^d = 0$, and find $Z(s^s) \propto I_0(s^s)$. The resulting strangeness condensate
is labeled by $t' = \infty$ in Figure~\ref{f:SU3}.

\section{Twisted quarks in fixed topology} \label{s:pert}

In this section, we discuss the effect of twisted boundary conditions  on the partition function in sectors of fixed
topological charge. If $Z_\nu$ corresponds to the partition function in a sector with topological charge $\nu$ then
$Z_\nu/Z$  can be interpreted as the probability of finding the gauge configuration with topological charge $\nu$. This can
be computed by introducing a $\theta$ term in the chiral Lagrangian and computing the Fourier transform of the partition
function with respect to $\theta$. Ref.~\cite{Leutwyler:1992yt} showed that the QCD partition function is
dominated by vanishing or small $\nu$ in the $\e$-regime.  We study whether this conclusion is modified in the presence of
twisted boundary conditions. For simplicity we consider only the $SU(2)_L \times SU(2)_R$ theory with $m_u = m_d$.

\begin{figure}
\epsfig{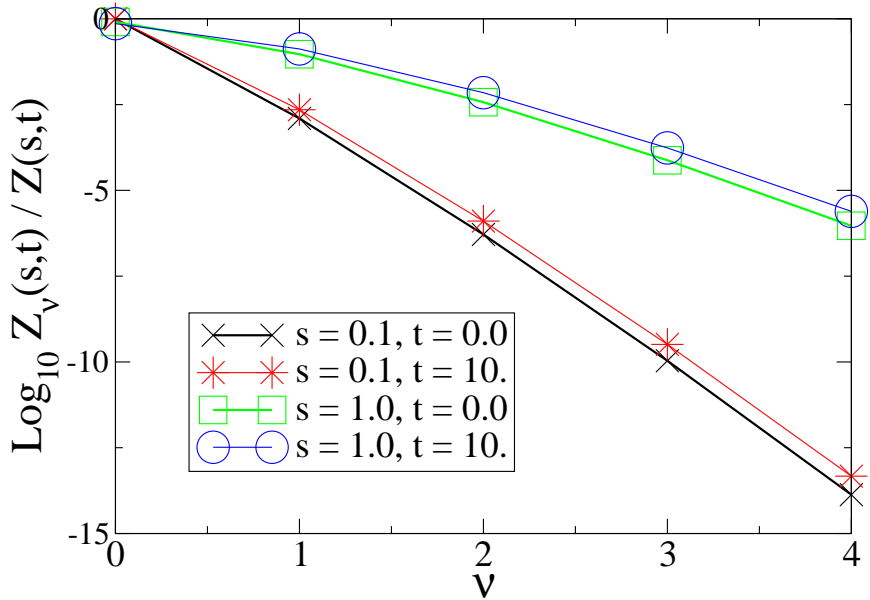}
\caption{The partition function for fixed topological charge. We plot ${\rm log}_{10} [Z_\nu (s,t) / Z (s,t)]$ as a function of $\nu$
for different $s$ and $t$. The solid lines are
provided merely to guide the eye.
}
\label{f:nu}
\end{figure}

In the twisted $\e$-regime, the partition function for vacuum angle $\th$ is given by
\begin{equation}
Z (\th)  = 
\int \cD \St_0 \, 
\exp 
\left\{ 
\frac{\beta V f_\pi^2}{4} 
\tr [B_\mu \St_0 B_\mu \St_0^\dagger 
+  
\Re \text{e} ( e^{i \th / 2}\chi \St_0 ) ]
\right\}
,\end{equation}
where we have specialized to the case of two flavors. In a sector with fixed topological
charge $\nu$, the corresponding partition function is
\begin{equation}
Z_\nu = \int_0^{2 \pi} \frac{d \th}{2 \pi} e^{- i \nu \th} Z(\th)
,\end{equation}
and can be straightforwardly evaluated. We find
\begin{equation}
Z_\nu (s,t) =
\frac{1}{\sqrt{\pi}} \int_0^{\pi} d\a \,
\frac{\sin \a}{\sqrt{t}} 
\Erfi ( \sqrt{t} \sin \a )
e^{- t \sin^2 \a}
I_{2 \nu}( 2 s \cos \a )
.\end{equation}
Note that $Z_\nu (s,t) =Z_{-\nu} (s,t)$. 
It is easy to show that in the limit $t \to 0$, we recover the result~\cite{Leutwyler:1992yt}
\bea 
Z_\nu(s,0)  = I_\nu(s)^2 - I_{\nu+1}(s) \,I_{\nu-1}(s) \, . \nn
\eea
In Figure~\ref{f:nu} the logarithm of the ratio $Z_\nu (s,t) / Z (s,t)$ is plotted for $\nu =0$
through $\nu = 4$. In this plot we show the result for two different values of $s$ with and without
twisted boundary conditions. We see that the dominance of the sector with zero topological charge in the $\e$-regime
is not modified by twisting. The numerical value of $Z_\nu (s,t) / Z (s,t)$ changes little between $t=0$ and $t=10$.

\section{Summary} \label{s:summy}

In this paper,  we have investigated the effects of twisted boundary conditions on the quark condensate in small volumes.
We worked in the twisted $\varepsilon$-regime of chiral perturbation theory  where $1/L \sim \e, \, B_\mu \sim m_\pi \sim
\e^2$ and all Goldstone boson zero modes are strongly coupled. The partition function and quark condensate were computed to leading order 
by performing the path integral over these zero modes. We also investigated the remnant $\e$-regime, where $1/L \sim  B_\mu \sim \e, \, m_\pi \sim \e^2$.
In this case only the flavor-neutral mesons are strongly coupled and the contribution from these modes dominates the partition
function.  It is the flavor-neutral zero modes that lead to a restoration of chiral symmetry at small volumes in the
remnant $\e$-regime.

The value of the quark condensate in small volumes affects the  masses of particles.
For example, the pion mass is given by
\begin{equation}
m_\pi^2 = \frac{\tr (\chi)}{2} \frac{\langle \ol q  q \rangle \phantom{8}}{\langle \ol q q \rangle_\infty} \,
,\end{equation}
and the masses of heavy particles such as the nucleon~\cite{Bedaque:2004dt}, pick up an additive renormalization,
\begin{equation}
m_N = m_0 - \a \, \tr (\chi) \frac{\langle \ol q q \rangle \phantom{8}}{\langle \ol q q \rangle_\infty} \,
.\end{equation}
When one imposes twisted boundary conditions on the quark fields, 
particle masses are modified since $\langle \bar q q \rangle$ 
is larger relative to its value in the absence of twisting.
Of course, the pion mass vanishes at very small volumes in any case.  

Lattice calculation of the dependence of quark condensates and particle masses on twisted boundary conditions
can be used to test the predictions of twisted chiral perturbation theory.  Moreover because the kinetic energy term
contributes to the twisted partition function,  it is in principle possible to determine both the infinite volume condensate and
the pion decay constant from numerically determining the finite volume condensate as a function of the twist angles and volume.  Lastly, we
showed that sectors with zero topological charge remain the dominant  contributors to the partition function in the
twisted and remnant $\e$-regimes.

For the results in this paper to be useful in actual simulations, it may be necessary to extend our calculations to include
effects  due to quenching or partial quenching as well as perturbative loop corrections from nonzero modes.  Note that the
simplicity of the zero-mode manifold in the remnant $\e$-regime may make it easier  to compute the partition
function for quenched and partially quenched theories with twisted quarks.  It would also be interesting to examine other
correlation functions, especially two point functions of currents. We hope this paper will stimulate further 
numerical and analytical work in this novel regime of QCD.

\begin{acknowledgments} B.C.T. would like to thank Will Detmold for
numerous discussions. B.C.T. and T.M. are supported in part by DOE grant 
DE-FG02-96ER40945. T.M. is also supported in part by DOE grant DE-AC05-84ER40150. 
 B.C.T. and T.M. thank the Institute for Nuclear Theory for their hospitality during the completion of this work. \end{acknowledgments}

\appendix

\bibliography{hb}

\begin{thebibliography}{49}
\expandafter\ifx\csname natexlab\endcsname\relax\def\natexlab#1{#1}\fi
\expandafter\ifx\csname bibnamefont\endcsname\relax
  \def\bibnamefont#1{#1}\fi
\expandafter\ifx\csname bibfnamefont\endcsname\relax
  \def\bibfnamefont#1{#1}\fi
\expandafter\ifx\csname citenamefont\endcsname\relax
  \def\citenamefont#1{#1}\fi
\expandafter\ifx\csname url\endcsname\relax
  \def\url#1{\texttt{#1}}\fi
\expandafter\ifx\csname urlprefix\endcsname\relax\def\urlprefix{URL }\fi
\providecommand{\bibinfo}[2]{#2}
\providecommand{\eprint}[2][]{\url{#2}}

\bibitem[{\citenamefont{Bernard et~al.}(2003)}]{Bernard:2002yk}
\bibinfo{author}{\bibfnamefont{C.}~\bibnamefont{Bernard}} \bibnamefont{et~al.},
  \bibinfo{journal}{Nucl. Phys. Proc. Suppl.} \textbf{\bibinfo{volume}{119}},
  \bibinfo{pages}{170} (\bibinfo{year}{2003}).

\bibitem[{\citenamefont{Gross and Kitazawa}(1982)}]{Gross:1982at}
\bibinfo{author}{\bibfnamefont{D.~J.} \bibnamefont{Gross}} \bibnamefont{and}
  \bibinfo{author}{\bibfnamefont{Y.}~\bibnamefont{Kitazawa}},
  \bibinfo{journal}{Nucl. Phys.} \textbf{\bibinfo{volume}{B206}},
  \bibinfo{pages}{440} (\bibinfo{year}{1982}).

\bibitem[{\citenamefont{Roberge and Weiss}(1986)}]{Roberge:1986mm}
\bibinfo{author}{\bibfnamefont{A.}~\bibnamefont{Roberge}} \bibnamefont{and}
  \bibinfo{author}{\bibfnamefont{N.}~\bibnamefont{Weiss}},
  \bibinfo{journal}{Nucl. Phys.} \textbf{\bibinfo{volume}{B275}},
  \bibinfo{pages}{734} (\bibinfo{year}{1986}).

\bibitem[{\citenamefont{Wiese}(1992)}]{Wiese:1991ku}
\bibinfo{author}{\bibfnamefont{U.~J.} \bibnamefont{Wiese}},
  \bibinfo{journal}{Nucl. Phys.} \textbf{\bibinfo{volume}{B375}},
  \bibinfo{pages}{45} (\bibinfo{year}{1992}).

\bibitem[{\citenamefont{Luscher et~al.}(1996)\citenamefont{Luscher, Sint,
  Sommer, and Weisz}}]{Luscher:1996sc}
\bibinfo{author}{\bibfnamefont{M.}~\bibnamefont{Luscher}},
  \bibinfo{author}{\bibfnamefont{S.}~\bibnamefont{Sint}},
  \bibinfo{author}{\bibfnamefont{R.}~\bibnamefont{Sommer}}, \bibnamefont{and}
  \bibinfo{author}{\bibfnamefont{P.}~\bibnamefont{Weisz}},
  \bibinfo{journal}{Nucl. Phys.} \textbf{\bibinfo{volume}{B478}},
  \bibinfo{pages}{365} (\bibinfo{year}{1996}).

\bibitem[{\citenamefont{Bucarelli et~al.}(1999)\citenamefont{Bucarelli,
  Palombi, Petronzio, and Shindler}}]{Bucarelli:1998mu}
\bibinfo{author}{\bibfnamefont{A.}~\bibnamefont{Bucarelli}},
  \bibinfo{author}{\bibfnamefont{F.}~\bibnamefont{Palombi}},
  \bibinfo{author}{\bibfnamefont{R.}~\bibnamefont{Petronzio}},
  \bibnamefont{and} \bibinfo{author}{\bibfnamefont{A.}~\bibnamefont{Shindler}},
  \bibinfo{journal}{Nucl. Phys.} \textbf{\bibinfo{volume}{B552}},
  \bibinfo{pages}{379} (\bibinfo{year}{1999}).

\bibitem[{\citenamefont{Kiskis et~al.}(2002)\citenamefont{Kiskis, Narayanan,
  and Neuberger}}]{Kiskis:2002gr}
\bibinfo{author}{\bibfnamefont{J.}~\bibnamefont{Kiskis}},
  \bibinfo{author}{\bibfnamefont{R.}~\bibnamefont{Narayanan}},
  \bibnamefont{and}
  \bibinfo{author}{\bibfnamefont{H.}~\bibnamefont{Neuberger}},
  \bibinfo{journal}{Phys. Rev.} \textbf{\bibinfo{volume}{D66}},
  \bibinfo{pages}{025019} (\bibinfo{year}{2002}).

\bibitem[{\citenamefont{Guagnelli et~al.}(2003)}]{Guagnelli:2003hw}
\bibinfo{author}{\bibfnamefont{M.}~\bibnamefont{Guagnelli}}
  \bibnamefont{et~al.} (\bibinfo{collaboration}{Zeuthen-Rome / ZeRo}),
  \bibinfo{journal}{Nucl. Phys.} \textbf{\bibinfo{volume}{B664}},
  \bibinfo{pages}{276} (\bibinfo{year}{2003}).

\bibitem[{\citenamefont{Kiskis et~al.}(2003)\citenamefont{Kiskis, Narayanan,
  and Neuberger}}]{Kiskis:2003rd}
\bibinfo{author}{\bibfnamefont{J.}~\bibnamefont{Kiskis}},
  \bibinfo{author}{\bibfnamefont{R.}~\bibnamefont{Narayanan}},
  \bibnamefont{and}
  \bibinfo{author}{\bibfnamefont{H.}~\bibnamefont{Neuberger}},
  \bibinfo{journal}{Phys. Lett.} \textbf{\bibinfo{volume}{B574}},
  \bibinfo{pages}{65} (\bibinfo{year}{2003}).

\bibitem[{\citenamefont{Kim and Christ}(2003)}]{Kim:2002np}
\bibinfo{author}{\bibfnamefont{C.-H.} \bibnamefont{Kim}} \bibnamefont{and}
  \bibinfo{author}{\bibfnamefont{N.~H.} \bibnamefont{Christ}},
  \bibinfo{journal}{Nucl. Phys. Proc. Suppl.} \textbf{\bibinfo{volume}{119}},
  \bibinfo{pages}{365} (\bibinfo{year}{2003}).

\bibitem[{\citenamefont{Kim}(2004)}]{Kim:2003xt}
\bibinfo{author}{\bibfnamefont{C.-H.} \bibnamefont{Kim}},
  \bibinfo{journal}{Nucl. Phys. Proc. Suppl.} \textbf{\bibinfo{volume}{129}},
  \bibinfo{pages}{197} (\bibinfo{year}{2004}).

\bibitem[{\citenamefont{Dagotto et~al.}(1990)\citenamefont{Dagotto, Moreo,
  Sugar, and Toussaint}}]{Dagotto:1989fw}
\bibinfo{author}{\bibfnamefont{E.}~\bibnamefont{Dagotto}},
  \bibinfo{author}{\bibfnamefont{A.}~\bibnamefont{Moreo}},
  \bibinfo{author}{\bibfnamefont{R.~L.} \bibnamefont{Sugar}}, \bibnamefont{and}
  \bibinfo{author}{\bibfnamefont{D.}~\bibnamefont{Toussaint}},
  \bibinfo{journal}{Phys. Rev.} \textbf{\bibinfo{volume}{B41}},
  \bibinfo{pages}{811} (\bibinfo{year}{1990}).

\bibitem[{\citenamefont{Hasenfratz and Toussaint}(1992)}]{Hasenfratz:1991ax}
\bibinfo{author}{\bibfnamefont{A.}~\bibnamefont{Hasenfratz}} \bibnamefont{and}
  \bibinfo{author}{\bibfnamefont{D.}~\bibnamefont{Toussaint}},
  \bibinfo{journal}{Nucl. Phys.} \textbf{\bibinfo{volume}{B371}},
  \bibinfo{pages}{539} (\bibinfo{year}{1992}).

\bibitem[{\citenamefont{Alford et~al.}(1999)\citenamefont{Alford, Kapustin, and
  Wilczek}}]{Alford:1998sd}
\bibinfo{author}{\bibfnamefont{M.~G.} \bibnamefont{Alford}},
  \bibinfo{author}{\bibfnamefont{A.}~\bibnamefont{Kapustin}}, \bibnamefont{and}
  \bibinfo{author}{\bibfnamefont{F.}~\bibnamefont{Wilczek}},
  \bibinfo{journal}{Phys. Rev.} \textbf{\bibinfo{volume}{D59}},
  \bibinfo{pages}{054502} (\bibinfo{year}{1999}).

\bibitem[{\citenamefont{Hart et~al.}(2001)\citenamefont{Hart, Laine, and
  Philipsen}}]{Hart:2000ef}
\bibinfo{author}{\bibfnamefont{A.}~\bibnamefont{Hart}},
  \bibinfo{author}{\bibfnamefont{M.}~\bibnamefont{Laine}}, \bibnamefont{and}
  \bibinfo{author}{\bibfnamefont{O.}~\bibnamefont{Philipsen}},
  \bibinfo{journal}{Phys. Lett.} \textbf{\bibinfo{volume}{B505}},
  \bibinfo{pages}{141} (\bibinfo{year}{2001}).

\bibitem[{\citenamefont{Allton et~al.}(2002)}]{Allton:2002zi}
\bibinfo{author}{\bibfnamefont{C.~R.} \bibnamefont{Allton}}
  \bibnamefont{et~al.}, \bibinfo{journal}{Phys. Rev.}
  \textbf{\bibinfo{volume}{D66}}, \bibinfo{pages}{074507}
  (\bibinfo{year}{2002}).

\bibitem[{\citenamefont{de~Forcrand and Philipsen}(2002)}]{deForcrand:2002ci}
\bibinfo{author}{\bibfnamefont{P.}~\bibnamefont{de~Forcrand}} \bibnamefont{and}
  \bibinfo{author}{\bibfnamefont{O.}~\bibnamefont{Philipsen}},
  \bibinfo{journal}{Nucl. Phys.} \textbf{\bibinfo{volume}{B642}},
  \bibinfo{pages}{290} (\bibinfo{year}{2002}).

\bibitem[{\citenamefont{D'Elia and Lombardo}(2003)}]{D'Elia:2002gd}
\bibinfo{author}{\bibfnamefont{M.}~\bibnamefont{D'Elia}} \bibnamefont{and}
  \bibinfo{author}{\bibfnamefont{M.-P.} \bibnamefont{Lombardo}},
  \bibinfo{journal}{Phys. Rev.} \textbf{\bibinfo{volume}{D67}},
  \bibinfo{pages}{014505} (\bibinfo{year}{2003}).

\bibitem[{\citenamefont{de~Forcrand and Philipsen}(2003)}]{deForcrand:2003hx}
\bibinfo{author}{\bibfnamefont{P.}~\bibnamefont{de~Forcrand}} \bibnamefont{and}
  \bibinfo{author}{\bibfnamefont{O.}~\bibnamefont{Philipsen}},
  \bibinfo{journal}{Nucl. Phys.} \textbf{\bibinfo{volume}{B673}},
  \bibinfo{pages}{170} (\bibinfo{year}{2003}).

\bibitem[{\citenamefont{Bedaque}(2004)}]{Bedaque:2004kc}
\bibinfo{author}{\bibfnamefont{P.~F.} \bibnamefont{Bedaque}},
  \bibinfo{journal}{Phys. Lett.} \textbf{\bibinfo{volume}{B593}},
  \bibinfo{pages}{82} (\bibinfo{year}{2004}).

\bibitem[{\citenamefont{de~Divitiis et~al.}(2004)\citenamefont{de~Divitiis,
  Petronzio, and Tantalo}}]{deDivitiis:2004kq}
\bibinfo{author}{\bibfnamefont{G.~M.} \bibnamefont{de~Divitiis}},
  \bibinfo{author}{\bibfnamefont{R.}~\bibnamefont{Petronzio}},
  \bibnamefont{and} \bibinfo{author}{\bibfnamefont{N.}~\bibnamefont{Tantalo}},
  \bibinfo{journal}{Phys. Lett.} \textbf{\bibinfo{volume}{B595}},
  \bibinfo{pages}{408} (\bibinfo{year}{2004}).

\bibitem[{\citenamefont{Sachrajda and Villadoro}(2005)}]{Sachrajda:2004mi}
\bibinfo{author}{\bibfnamefont{C.~T.} \bibnamefont{Sachrajda}}
  \bibnamefont{and}
  \bibinfo{author}{\bibfnamefont{G.}~\bibnamefont{Villadoro}},
  \bibinfo{journal}{Phys. Lett.} \textbf{\bibinfo{volume}{B609}},
  \bibinfo{pages}{73} (\bibinfo{year}{2005}).

\bibitem[{\citenamefont{Bedaque and Chen}(2004)}]{Bedaque:2004ax}
\bibinfo{author}{\bibfnamefont{P.~F.} \bibnamefont{Bedaque}} \bibnamefont{and}
  \bibinfo{author}{\bibfnamefont{J.-W.} \bibnamefont{Chen}}
  (\bibinfo{year}{2004}), \eprint{hep-lat/0412023}.

\bibitem[{\citenamefont{Tiburzi}(2005)}]{Tiburzi:2005hg}
\bibinfo{author}{\bibfnamefont{B.~C.} \bibnamefont{Tiburzi}}
  (\bibinfo{year}{2005}), \eprint{hep-lat/0504002}.

\bibitem[{\citenamefont{Jolicoeur and Morel}(1985)}]{Jolicoeur:1985iw}
\bibinfo{author}{\bibfnamefont{T.}~\bibnamefont{Jolicoeur}} \bibnamefont{and}
  \bibinfo{author}{\bibfnamefont{A.}~\bibnamefont{Morel}},
  \bibinfo{journal}{Nucl. Phys.} \textbf{\bibinfo{volume}{B262}},
  \bibinfo{pages}{627} (\bibinfo{year}{1985}).

\bibitem[{\citenamefont{Leutwyler}(1987)}]{Leutwyler:1987ak}
\bibinfo{author}{\bibfnamefont{H.}~\bibnamefont{Leutwyler}},
  \bibinfo{journal}{Phys. Lett.} \textbf{\bibinfo{volume}{B189}},
  \bibinfo{pages}{197} (\bibinfo{year}{1987}).

\bibitem[{\citenamefont{Gasser and Leutwyler}(1987)}]{Gasser:1987ah}
\bibinfo{author}{\bibfnamefont{J.}~\bibnamefont{Gasser}} \bibnamefont{and}
  \bibinfo{author}{\bibfnamefont{H.}~\bibnamefont{Leutwyler}},
  \bibinfo{journal}{Phys. Lett.} \textbf{\bibinfo{volume}{B188}},
  \bibinfo{pages}{477} (\bibinfo{year}{1987}).

\bibitem[{\citenamefont{Gasser and Leutwyler}(1988)}]{Gasser:1987zq}
\bibinfo{author}{\bibfnamefont{J.}~\bibnamefont{Gasser}} \bibnamefont{and}
  \bibinfo{author}{\bibfnamefont{H.}~\bibnamefont{Leutwyler}},
  \bibinfo{journal}{Nucl. Phys.} \textbf{\bibinfo{volume}{B307}},
  \bibinfo{pages}{763} (\bibinfo{year}{1988}).

\bibitem[{\citenamefont{Hasenfratz and Leutwyler}(1990)}]{Hasenfratz:1989pk}
\bibinfo{author}{\bibfnamefont{P.}~\bibnamefont{Hasenfratz}} \bibnamefont{and}
  \bibinfo{author}{\bibfnamefont{H.}~\bibnamefont{Leutwyler}},
  \bibinfo{journal}{Nucl. Phys.} \textbf{\bibinfo{volume}{B343}},
  \bibinfo{pages}{241} (\bibinfo{year}{1990}).

\bibitem[{\citenamefont{Hansen}(1990)}]{Hansen:1990un}
\bibinfo{author}{\bibfnamefont{F.~C.} \bibnamefont{Hansen}},
  \bibinfo{journal}{Nucl. Phys.} \textbf{\bibinfo{volume}{B345}},
  \bibinfo{pages}{685} (\bibinfo{year}{1990}).

\bibitem[{\citenamefont{Hansen and Leutwyler}(1991)}]{Hansen:1990yg}
\bibinfo{author}{\bibfnamefont{F.~C.} \bibnamefont{Hansen}} \bibnamefont{and}
  \bibinfo{author}{\bibfnamefont{H.}~\bibnamefont{Leutwyler}},
  \bibinfo{journal}{Nucl. Phys.} \textbf{\bibinfo{volume}{B350}},
  \bibinfo{pages}{201} (\bibinfo{year}{1991}).

\bibitem[{\citenamefont{Jackson et~al.}(1996)\citenamefont{Jackson, Sener, and
  Verbaarschot}}]{Jackson:1996jb}
\bibinfo{author}{\bibfnamefont{A.~D.} \bibnamefont{Jackson}},
  \bibinfo{author}{\bibfnamefont{M.~K.} \bibnamefont{Sener}}, \bibnamefont{and}
  \bibinfo{author}{\bibfnamefont{J.~J.~M.} \bibnamefont{Verbaarschot}},
  \bibinfo{journal}{Phys. Lett.} \textbf{\bibinfo{volume}{B387}},
  \bibinfo{pages}{355} (\bibinfo{year}{1996}).

\bibitem[{\citenamefont{Osborn et~al.}(1999)\citenamefont{Osborn, Toublan, and
  Verbaarschot}}]{Osborn:1998qb}
\bibinfo{author}{\bibfnamefont{J.~C.} \bibnamefont{Osborn}},
  \bibinfo{author}{\bibfnamefont{D.}~\bibnamefont{Toublan}}, \bibnamefont{and}
  \bibinfo{author}{\bibfnamefont{J.~J.~M.} \bibnamefont{Verbaarschot}},
  \bibinfo{journal}{Nucl. Phys.} \textbf{\bibinfo{volume}{B540}},
  \bibinfo{pages}{317} (\bibinfo{year}{1999}).

\bibitem[{\citenamefont{Damgaard et~al.}(1999)\citenamefont{Damgaard, Osborn,
  Toublan, and Verbaarschot}}]{Damgaard:1998xy}
\bibinfo{author}{\bibfnamefont{P.~H.} \bibnamefont{Damgaard}},
  \bibinfo{author}{\bibfnamefont{J.~C.} \bibnamefont{Osborn}},
  \bibinfo{author}{\bibfnamefont{D.}~\bibnamefont{Toublan}}, \bibnamefont{and}
  \bibinfo{author}{\bibfnamefont{J.~J.~M.} \bibnamefont{Verbaarschot}},
  \bibinfo{journal}{Nucl. Phys.} \textbf{\bibinfo{volume}{B547}},
  \bibinfo{pages}{305} (\bibinfo{year}{1999}).

\bibitem[{\citenamefont{Damgaard et~al.}(2000)\citenamefont{Damgaard, Edwards,
  Heller, and Narayanan}}]{Damgaard:1999tk}
\bibinfo{author}{\bibfnamefont{P.~H.} \bibnamefont{Damgaard}},
  \bibinfo{author}{\bibfnamefont{R.~G.} \bibnamefont{Edwards}},
  \bibinfo{author}{\bibfnamefont{U.~M.} \bibnamefont{Heller}},
  \bibnamefont{and}
  \bibinfo{author}{\bibfnamefont{R.}~\bibnamefont{Narayanan}},
  \bibinfo{journal}{Phys. Rev.} \textbf{\bibinfo{volume}{D61}},
  \bibinfo{pages}{094503} (\bibinfo{year}{2000}).

\bibitem[{\citenamefont{Hernandez et~al.}(1999)\citenamefont{Hernandez, Jansen,
  and Lellouch}}]{Hernandez:1999cu}
\bibinfo{author}{\bibfnamefont{P.}~\bibnamefont{Hernandez}},
  \bibinfo{author}{\bibfnamefont{K.}~\bibnamefont{Jansen}}, \bibnamefont{and}
  \bibinfo{author}{\bibfnamefont{L.}~\bibnamefont{Lellouch}},
  \bibinfo{journal}{Phys. Lett.} \textbf{\bibinfo{volume}{B469}},
  \bibinfo{pages}{198} (\bibinfo{year}{1999}).

\bibitem[{\citenamefont{Damgaard et~al.}(2002)\citenamefont{Damgaard,
  Diamantini, Hernandez, and Jansen}}]{Damgaard:2001js}
\bibinfo{author}{\bibfnamefont{P.~H.} \bibnamefont{Damgaard}},
  \bibinfo{author}{\bibfnamefont{M.~C.} \bibnamefont{Diamantini}},
  \bibinfo{author}{\bibfnamefont{P.}~\bibnamefont{Hernandez}},
  \bibnamefont{and} \bibinfo{author}{\bibfnamefont{K.}~\bibnamefont{Jansen}},
  \bibinfo{journal}{Nucl. Phys.} \textbf{\bibinfo{volume}{B629}},
  \bibinfo{pages}{445} (\bibinfo{year}{2002}).

\bibitem[{\citenamefont{Damgaard et~al.}(2003)\citenamefont{Damgaard,
  Hernandez, Jansen, Laine, and Lellouch}}]{Damgaard:2002qe}
\bibinfo{author}{\bibfnamefont{P.~H.} \bibnamefont{Damgaard}},
  \bibinfo{author}{\bibfnamefont{P.}~\bibnamefont{Hernandez}},
  \bibinfo{author}{\bibfnamefont{K.}~\bibnamefont{Jansen}},
  \bibinfo{author}{\bibfnamefont{M.}~\bibnamefont{Laine}}, \bibnamefont{and}
  \bibinfo{author}{\bibfnamefont{L.}~\bibnamefont{Lellouch}},
  \bibinfo{journal}{Nucl. Phys.} \textbf{\bibinfo{volume}{B656}},
  \bibinfo{pages}{226} (\bibinfo{year}{2003}).

\bibitem[{\citenamefont{Bietenholz et~al.}(2004)\citenamefont{Bietenholz,
  Chiarappa, Jansen, Nagai, and Shcheredin}}]{Bietenholz:2003bj}
\bibinfo{author}{\bibfnamefont{W.}~\bibnamefont{Bietenholz}},
  \bibinfo{author}{\bibfnamefont{T.}~\bibnamefont{Chiarappa}},
  \bibinfo{author}{\bibfnamefont{K.}~\bibnamefont{Jansen}},
  \bibinfo{author}{\bibfnamefont{K.~I.} \bibnamefont{Nagai}}, \bibnamefont{and}
  \bibinfo{author}{\bibfnamefont{S.}~\bibnamefont{Shcheredin}},
  \bibinfo{journal}{JHEP} \textbf{\bibinfo{volume}{02}}, \bibinfo{pages}{023}
  (\bibinfo{year}{2004}).

\bibitem[{\citenamefont{Giusti et~al.}(2004)\citenamefont{Giusti, Hernandez,
  Laine, Weisz, and Wittig}}]{Giusti:2004yp}
\bibinfo{author}{\bibfnamefont{L.}~\bibnamefont{Giusti}},
  \bibinfo{author}{\bibfnamefont{P.}~\bibnamefont{Hernandez}},
  \bibinfo{author}{\bibfnamefont{M.}~\bibnamefont{Laine}},
  \bibinfo{author}{\bibfnamefont{P.}~\bibnamefont{Weisz}}, \bibnamefont{and}
  \bibinfo{author}{\bibfnamefont{H.}~\bibnamefont{Wittig}},
  \bibinfo{journal}{JHEP} \textbf{\bibinfo{volume}{04}}, \bibinfo{pages}{013}
  (\bibinfo{year}{2004}).

\bibitem[{\citenamefont{Fukaya et~al.}(2005)\citenamefont{Fukaya, Hashimoto,
  and Ogawa}}]{Fukaya:2005yg}
\bibinfo{author}{\bibfnamefont{H.}~\bibnamefont{Fukaya}},
  \bibinfo{author}{\bibfnamefont{S.}~\bibnamefont{Hashimoto}},
  \bibnamefont{and} \bibinfo{author}{\bibfnamefont{K.}~\bibnamefont{Ogawa}}
  (\bibinfo{year}{2005}), \eprint{hep-lat/0504018}.

\bibitem[{\citenamefont{Leutwyler and Smilga}(1992)}]{Leutwyler:1992yt}
\bibinfo{author}{\bibfnamefont{H.}~\bibnamefont{Leutwyler}} \bibnamefont{and}
  \bibinfo{author}{\bibfnamefont{A.}~\bibnamefont{Smilga}},
  \bibinfo{journal}{Phys. Rev.} \textbf{\bibinfo{volume}{D46}},
  \bibinfo{pages}{5607} (\bibinfo{year}{1992}).

\bibitem[{\citenamefont{Toublan and Verbaarschot}(2001)}]{Toublan:1999hx}
\bibinfo{author}{\bibfnamefont{D.}~\bibnamefont{Toublan}} \bibnamefont{and}
  \bibinfo{author}{\bibfnamefont{J.~J.~M.} \bibnamefont{Verbaarschot}},
  \bibinfo{journal}{Int. J. Mod. Phys.} \textbf{\bibinfo{volume}{B15}},
  \bibinfo{pages}{1404} (\bibinfo{year}{2001}).

\bibitem[{\citenamefont{Splittorff and Verbaarschot}(2004)}]{Splittorff:2003cu}
\bibinfo{author}{\bibfnamefont{K.}~\bibnamefont{Splittorff}} \bibnamefont{and}
  \bibinfo{author}{\bibfnamefont{J.~J.~M.} \bibnamefont{Verbaarschot}},
  \bibinfo{journal}{Nucl. Phys.} \textbf{\bibinfo{volume}{B683}},
  \bibinfo{pages}{467} (\bibinfo{year}{2004}).

\bibitem[{\citenamefont{Akemann et~al.}(2005)\citenamefont{Akemann, Osborn,
  Splittorff, and Verbaarschot}}]{Akemann:2004dr}
\bibinfo{author}{\bibfnamefont{G.}~\bibnamefont{Akemann}},
  \bibinfo{author}{\bibfnamefont{J.~C.} \bibnamefont{Osborn}},
  \bibinfo{author}{\bibfnamefont{K.}~\bibnamefont{Splittorff}},
  \bibnamefont{and} \bibinfo{author}{\bibfnamefont{J.~J.~M.}
  \bibnamefont{Verbaarschot}}, \bibinfo{journal}{Nucl. Phys.}
  \textbf{\bibinfo{volume}{B712}}, \bibinfo{pages}{287} (\bibinfo{year}{2005}).

\bibitem[{\citenamefont{Byrd and Sudarshan}(1998)}]{Byrd:1998uw}
\bibinfo{author}{\bibfnamefont{M.~S.} \bibnamefont{Byrd}} \bibnamefont{and}
  \bibinfo{author}{\bibfnamefont{E.~C.~G.} \bibnamefont{Sudarshan}},
  \bibinfo{journal}{J. Phys.} \textbf{\bibinfo{volume}{A31}},
  \bibinfo{pages}{9255} (\bibinfo{year}{1998}).

\bibitem[{\citenamefont{Byrd}(1998)}]{Byrd:1998??}
\bibinfo{author}{\bibfnamefont{M.~S.} \bibnamefont{Byrd}}, \bibinfo{journal}{J.
  Math. Phys.} \textbf{\bibinfo{volume}{39}}, \bibinfo{pages}{11}
  (\bibinfo{year}{1998}).

\bibitem[{\citenamefont{Holland}(1969)}]{Holland:1969??}
\bibinfo{author}{\bibfnamefont{D.~F.} \bibnamefont{Holland}},
  \bibinfo{journal}{J. Math. Phys.} \textbf{\bibinfo{volume}{10}},
  \bibinfo{pages}{1903} (\bibinfo{year}{1969}).

\bibitem[{\citenamefont{Bedaque et~al.}(2005)\citenamefont{Bedaque,
  Griesshammer, and Rupak}}]{Bedaque:2004dt}
\bibinfo{author}{\bibfnamefont{P.~F.} \bibnamefont{Bedaque}},
  \bibinfo{author}{\bibfnamefont{H.~W.} \bibnamefont{Griesshammer}},
  \bibnamefont{and} \bibinfo{author}{\bibfnamefont{G.}~\bibnamefont{Rupak}},
  \bibinfo{journal}{Phys. Rev.} \textbf{\bibinfo{volume}{D71}},
  \bibinfo{pages}{054015} (\bibinfo{year}{2005}).

\end{thebibliography}

\end{document}